\documentclass[11pt,english]{article}
\usepackage[a4paper,top=2cm,bottom=2cm,left=2cm,right=2cm]{geometry}

\usepackage{mathptmx}
\usepackage[english]{babel}
\usepackage[utf8x]{inputenc}
\usepackage[T1]{fontenc}
\usepackage{ae,aecompl}
\usepackage{amsfonts}
\usepackage{amssymb}
\usepackage[superscript,biblabel]{cite}
\usepackage[numbers]{natbib}

\usepackage[fleqn]{amsmath}
\usepackage{graphicx}
\usepackage{subcaption}
\usepackage{url}
\usepackage[colorlinks=true, allcolors=blue]{hyperref}
\usepackage{authblk}
\begin{document}
\title{\textbf{Laser propagation in a highly magnetized over-dense plasma}}

\author[1]{K.Li\thanks{Corresponding Author: 93793444@qq.com}}
\author[2]{W.Yu}
\affil[1]{Institute of Physics of the ASCR, ELI-Beamlines, Na Slovance 2, 18221 Prague, Czech Republic}
\affil[2]{Shanghai Institute of Optics and Fine Mechanics, Chinese Academy of Science, Shanghai 201800, China}

\date{} % hide date with blank
\renewcommand\Affilfont{\itshape\small}

\maketitle
\vspace*{-1cm}
             
% remove title of "abstract"
\renewcommand{\abstractname}{\vspace{-\baselineskip}}

\begin{abstract}
\noindent Propagation of right-hand circularly polarized laser into highly magnetized over-dense collisional plasma is analytically studied from basic equations. Highly magnetized plasma with density of $10^{23}$ $cm^{-3}$ is efficiently heated from 100 eV to 1 keV within depth of around 20 $\mu{m}$ in one nanosecond by Nd:YAG laser with moderate irradiance of around $10^{14}$ $Wcm^{-2}$. 
\end{abstract}

%\keywords{magnetized plasma, circularly polarized laser, laser fusion}
%\pacs{52.00.00}% PACS, the Physics and Astronomy

\section{INTRODUCTION}
The research on inertial confinement fusion started as early as 1960s. In this process, the energy of incident laser is mainly deposited in the corona region of laser-produced plasma and energy transfer to compressed over-dense deuterium-tritium fuel is achieved through shock waves or secondary particles such as energetic electrons or ions \cite{atzeni2004physics,tabak2014alternative}, which intrinsically decreases the conversion efficiency from laser to compressed fuel. The interaction of  electromagnetic (EM) wave with magnetized plasma has been studied since 1960s and described in several popular textbooks \cite{ginzburg1970propagation,chen1985introduction,Boyd}, where early study was mainly in the region of weak magnetic field and low frequency EM wave, for example, at Earth's magnetic field and radio frequency \cite{chen1985introduction}. In our previous work \cite{li2019propagation}, the interaction of circularly polarized (CP) laser at moderate irradiance with highly magnetized over-dense plasma was preliminarily studied, where laser could propagate into overdense plasma and its basic physics is described briefly below. 

When left-hand circularly polarized (LHCP) and right-hand circularly polarized (RHCP) laser propagate in magnetized plasma with propagation direction parallel to that of magnetic field, the relative permittivity is shown below, respectively.
\begin{align}
& \epsilon_{L}=1-n/(1+B)\\
& \epsilon_{R}=1-n/(1-B)=1+n/(B-1)
\end{align}
It is clear that LHCP laser is reflected at density where $n=1+B$ in both weakly ($B<1$) or highly ($B>1$) magnetized plasma with $0<\epsilon_L<1$. RHCP laser is also confined in regime of $0<\epsilon_R<1$ when $B<1$. However when $B>1$, it is found that the relative permittivity is always larger than 1, $\epsilon_R>1$, which means that RHCP can propagate into "any" over-dense magnetized plasma.  Here, $n=n_e/n_c$ is dimensionless plasma density and $B=B_0/B_c$ is dimensionless magnetic field, where $n_e$ is electron plasma density, $n_c$ is critical density of plasma with $n_c =m_e\omega_L^2/4\pi e^2 \approx 1.12\times 10^{21}\lambda_{L,\mu{m}}^{-2}$ $cm^{-3}$, $B_0$ is amplitude of extermal magnetic field, $B_c$ is amplitude of magnetic field related with electron gyrofrequency with $B_c = m_ec\omega_L/e \approx 1.07\times 10^4\lambda_{L,\mu{m}}^{-1}$ tesla, $\omega_L$ is angular frequency of laser in unit of radian per second and $\lambda_L$ is laser wavelength in unit of micrometer. To satisfy the condition that $B>1$, the amplitude of magnetic field must be extremely high for lasers, for example, $B_0>10^4$ tesla for fundamental wavelength of Nd:YAG laser.

Neodymium magnet is the most widely used static magnet with amplitude of magnetic field being only a couple of tesla. Howeve, the amplitude of pulsed magnetic field is much stronger and has been increased greatly in recent years thanks to various new techniques and physics \cite{battesti2018high,knauer2010compressing,fujioka2013kilotesla}, such as superconductor, laser-driven capacity coil, flux compression, interaction of intense laser with plasma and et al. For example, microsecond magnetic field of 1200 T is generated by electromagnetic flux-compression \cite{nakamura2018record} and picosecond magnetic field as high as $7\times 10^4$ tesla is generated by intense laser during its interaction with solid target \cite{wagner2004laboratory}. As a result, application of RHCP laser in overdense plasma in region of $B > 1$ might become reality in the not-too-far future. In previous work \cite{li2019propagation}, the temperature of plasma was assumed to be constant. In this paper, the plasma heating will analytically studied with process of collisional absorption because it is the dominant process for laser irradiance below $10^{14}$ $cm^{-2}$.  Appropriate parameters will also be investigated as to heating magnetized overdense cold plasma to temperature of 10s eV to 1000s eV which is of great interest in fields such as warm dense matter, fusion and et al.

\section{LASER PROPAGATION IN HIGHLY MAGNETIZED PLASMA}
Consider a step-like density profile with vacuum in region of $z<0$ and over-dense plasma in region of $z\geq 0$, where external magnetic field is applied along z+ direction. Circularly polarized (CP) laser propagates from vacuum side and normally onto plasma with direction parallel to that of external magnetic field, where laser irradiance is only slightly above the ionization threshold excluding both pondermotive and relativistic effects. In this way, the governing equations of the laser interactions with magnetized plasma are written as 
\begin{align}
\begin{split}
&\partial_t\textbf{u} =-\textbf{E}-\textbf{u}\times\textbf{B}_z-\nu\textbf{u}-n^{-1}\nabla{p} \\
&\nabla\times\nabla\times\textbf{E}+\partial_{tt}\textbf{E}=\partial_t(n\textbf{u})\\
&\nabla\cdot\textbf{E}=Zn_i-n\\
&\partial_tn=-\nabla\cdot(n\textbf{u})
\end{split}
\end{align}
where $p=nT$ is dimensionless plasma pressure, $T=T_{eV}/(m_ec^2)$ is dimensionless plasma temperature with $T_{eV}$ being plasma temperature in unit of electron volt, $m_e$ being static mass of electron and $c$ being velocity of light in vacuum, $E=E_L/(m_ec\omega_L/e)$ is dimensionless electric field of laser, $u=u_e/c$ is dimensionless velocity, $x=k_Lx_0$ is dimensionless space and $t=\omega_L t_0$ is dimensionless time, respectively. And $\nu$ is the electron-ion collisional rate normalized to angular frequency of laser
\begin{align} {\label{equ:collisionRate}}
&\nu=\nu_{ei}/\omega_L\approx 1.72 ln\Lambda Z n \lambda_{L}^{-1}T_{eV}^{-3/2}
\end{align}
where $\nu_{ei}=2.91\times 10^{-6} Zn_e ln\Lambda T_{eV}^{-3/2}s^{-1}$\cite{kruer2003physics} is electron-ion collisional rate,  $Z$ is number of free electron per atom set to 1 and $ln\Lambda$ is Coulomb logarithm set to be 1 here.  

As a vector, the electric field in plasma can be written in the form of $\textbf{E}=\textbf{E}_t+\textbf{E}_l$, with transverse components being $\textbf{E}_t=-\partial_t\textbf{a}$, $\nabla \cdot \textbf{E}_t=0$ and longitudinal component being $\textbf{E}_l=-\nabla\phi$, $\nabla \times \textbf{E}_l=0$, respectively. In our case where laser is normally incident into plasma, the transverse mode $\textbf{E}_t$ is perpendicular to laser direction, while the longitudinal component $\textbf{E}_l$ parallel to it. The inter-conversion between the transverse and longitudinal modes occurs when laser frequency is close to the local plasma frequency, however the plasma density under consideration here is much higher than the critical density of laser, thus the mode conversion (or resonance) between the transverse and longitudinal modes is not likely to occur. In this way, the starting equations can be simplified to be
\begin{align}
\begin{split}
&\partial_t (\textbf{u}_t-\textbf{a})+\nu \textbf{u}_t + \textbf{u}_t\times\textbf{B}_z=-\partial_t \textbf{u}_l-\textbf{E}_L-\nu\textbf{u}_l-\nabla{p}/n=0\\
& \nabla^2\textbf{a} -\partial_{tt}\textbf{a}-n\textbf{u}_t= n\textbf{u}_l-\partial_t\textbf{E}_L=0
\end{split}
\end{align}

\section{TRANSEVERSE COMPONENT} \label{sec3} 
Because fusion laser usually has temporal and spatial flat-top profile, the dimensionless potential vector of laser is simply expressed as below within the temporal and spatial range of laser pulse
\begin{align}\label{eq:1}
& \textbf{a}=a_0e^{i(k_{\sigma}z-t)}(\vec{x}-i\sigma\vec{y})
\end{align}
where $k_{\sigma}=\sqrt{\epsilon_{\sigma}}=k_{\sigma{r}}+ik_{\sigma{i}}$ is the dimensionless complex wave vector with $\epsilon_{\sigma}=\epsilon_{\sigma{r}}+i\epsilon_{\sigma{i}}$ is relative permittivity of plasma, $\sigma = -1$ for RHCP and $\sigma=1$ for LHCP laser, respectively. The dimensionless velocity of electron in CP laser is expressed as $\textbf{u}_t=u_t(\vec{x}-i\sigma\vec{y})$, thus the force equation for transverse movement of electron in the field of CP laser becomes
\begin{align}
&\partial_t \textbf{u}_t =\partial_t\textbf{a}-\textbf{u}_t\times\textbf{B}_z-\nu\textbf{u}_t\\
& \textbf{u}_t=\frac{\textbf{a}}{1+\sigma B_z+i\nu}
\end{align}
Then relative permittivity of magnetized plasma is obtained here, same as in previous work \cite{li2019propagation}. 
\begin{align}{\label{equ:permittivity}}
& \epsilon_{\sigma}=1-\frac{n}{1+\sigma B_z+i\nu}
\end{align}
The reflection and transmission coefficients of laser at vacuum-plasma interface are given by Fresnel equations \cite{born1980principle}
\begin{align}{\label{equ:heating}}
& r_{\sigma}=(1-k_{\sigma})/(1+k_{\sigma})\ \ \ \ \ t_{\sigma}=2/(1+k_{\sigma})
\end{align}
As laser propagates inside plasma, it deposits energy obeying equation of energy conservation \cite{jackson1999classical}
\begin{align}{\label{equ:heating}}
& \partial_t{(w_{P}+w_{L})}=-\nabla\cdot \textbf{S}_t-Im(\textbf{J}_t\cdot \textbf{E}_t)
\end{align}
Here, $w_P=3/2\cdot nT$ is dimensionless energy density of plasma, assuming that plasma behaves like ideal gas \cite{paul2006high}. $w_L=(\epsilon_rE^2+B^2)/2=(1+\epsilon_r)a^2/2$ is dimensionless energy density of laser in plasma. $\nabla\cdot \textbf{S}_t=\partial_z (a_1a_1^*)$ is laser absorption rate in plasma where $a_1=t_{\sigma}\cdot a$ is dimensionless potential vector of transmitted laser. Imaginary part of  $-\textbf{J}_t\cdot \textbf{E}_t=-nu_t\cdot a_1$ is the plasma heating rate due to transverse electron current. After required parameters are inserted into equ. \ref{equ:heating}, plasma temperature at vacuum-plasma interface, T(0, t), is obtained and transmission coefficients of laser into plasma is calculated. Then equ.\ref{equ:heating} is applied again in region of $z>0$ to obtain temperature distribution of plasma, T(z, t). Thus the propagation of CP laser into plasma is fully understood.

Before solving the above equations, the permittivity and absorption rate are discussed in case that magnetic field is extremely high ($B\gg 1$) and plasma temperature is either very low ($\nu\gg B$) or very high ($\nu\ll B$), as shown below. 
\begin{align}{\label{equ: approximation}}
&\epsilon_R \approx 1+\frac{n/B}{1+\nu^2/B^2}+i\frac{n/B}{1+\nu^2/B^2}\frac{\nu}{B}\approx  \left\{ \begin{array}{ll}
1+nB/\nu^2+in/\nu & \textrm{$\nu \gg B$}\\
1+n/B+in\nu/B^2 & \textrm{$\nu \ll B$}
\end{array} \right.
\end{align}
When plasma temperature is very low ($\nu\gg B$), laser absorption rate increases with plasma temperature or laser heating based on relationship between dimensionless collisional rate and plasma temperature shown in equ. \ref{equ:collisionRate}.  When plasma temperature is very high ($\nu\ll B$), laser absorption decreases with higher temperature and even becomes negligible at extremely high temperature, i.e., $\epsilon_R\approx 1+n/B$. The threshold of plasma temperature in case of $\nu\approx B$ is $T_{eV}\approx (1.72ln\Lambda \lambda_L^{-1}Zn/B)^{2/3}$, which is around 7 eV with parameters of $B=10$, $n=100$, Z=1 and $\lambda_L=1.06$ $\mu{m}$ as applied in next section.

\section{RESULTS AND DISCUSSIONS}

\begin{figure*}[hbt!]
  \centering
   \begin{subfigure}[b]{0.49\linewidth}
    \includegraphics[width=\linewidth]{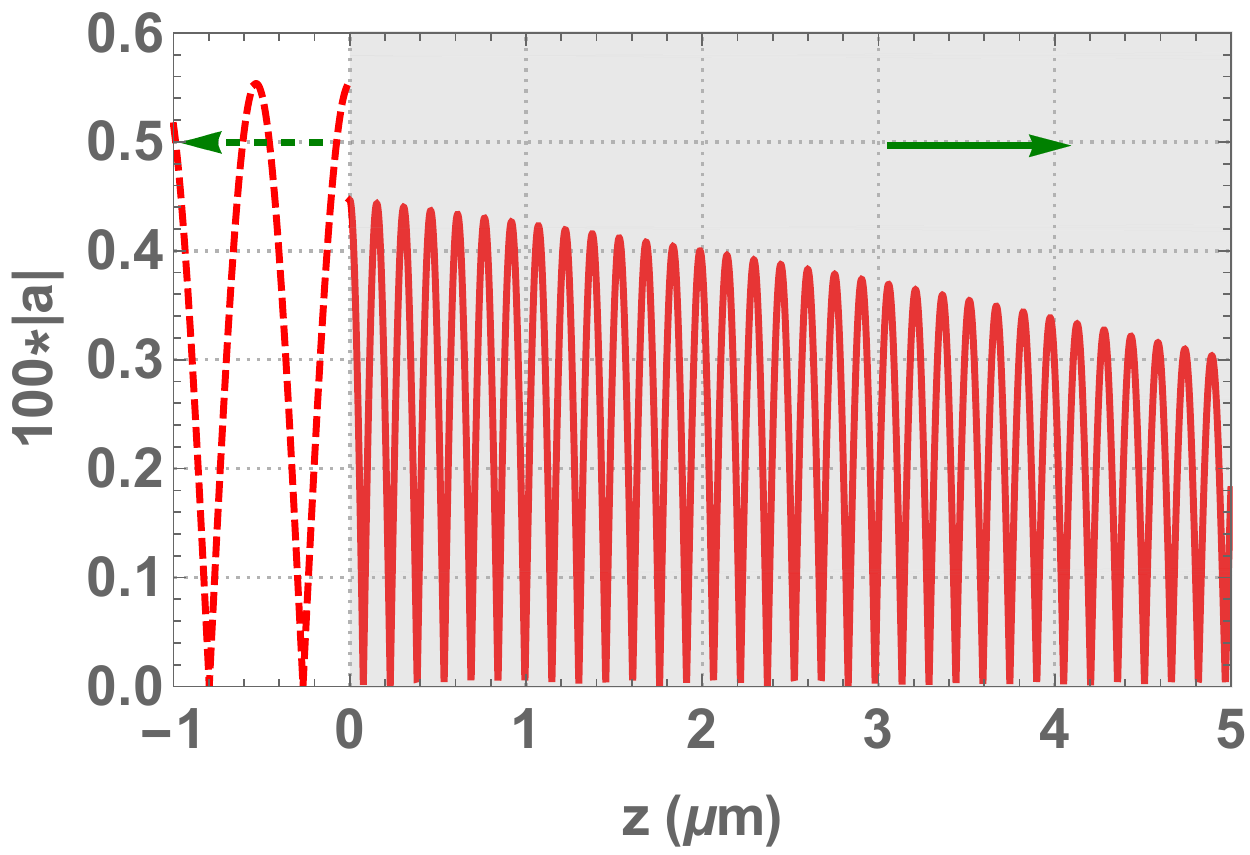}
    \caption{\label{fig:Ef}}
  \end{subfigure}
  \begin{subfigure}[b]{0.5\linewidth}
    \includegraphics[width=\linewidth]{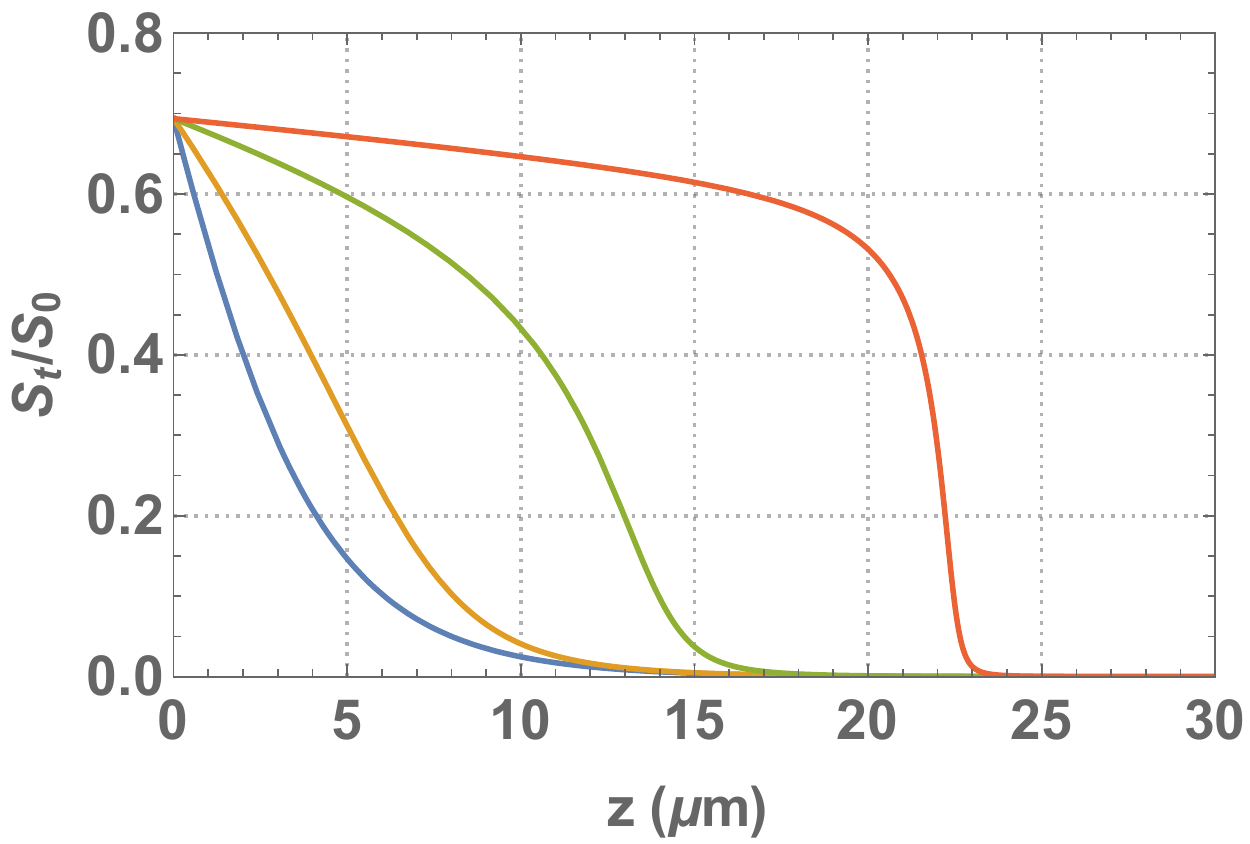}
    \caption{\label{fig:Temp} }
  \end{subfigure}
  \captionsetup{justification=raggedright,singlelinecheck=false}
  \caption{\label{fig:Efield} (a) Amplitude of electric fields for reflected (dashed) and transmitted (solid) RHCP laser at $t=10$ ps, where plasma is in region of z > 0 and vacuum in region of z < 0; (b) Dimensionless poynting vector of RHCP laser in plasma is shown at t = 1 ps, 10 ps, 100 ps and 1 ns (from left to right), respectively. Parameters: $a_0=0.01$, $\lambda_L=1.06$ $\mu{m}$, $B_z=10$, $n=100$ and initial plasma temperature is $T_0$ = 100 eV.}
\end{figure*}

\begin{figure*}[hbt!]
  \centering
   \begin{subfigure}[b]{0.51\linewidth}
    \includegraphics[width=\linewidth]{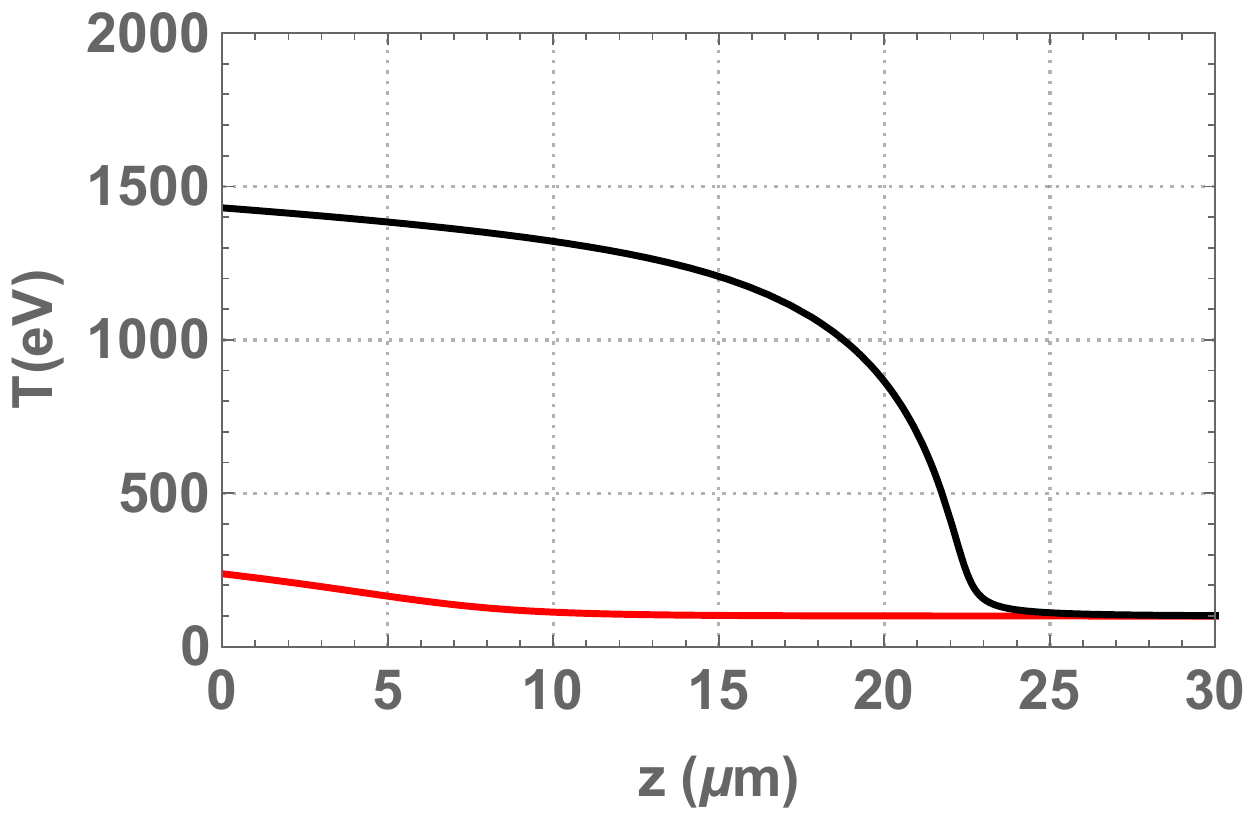}
    \caption{\label{fig:S}}
  \end{subfigure}
  \begin{subfigure}[b]{0.48\linewidth}
    \includegraphics[width=\linewidth]{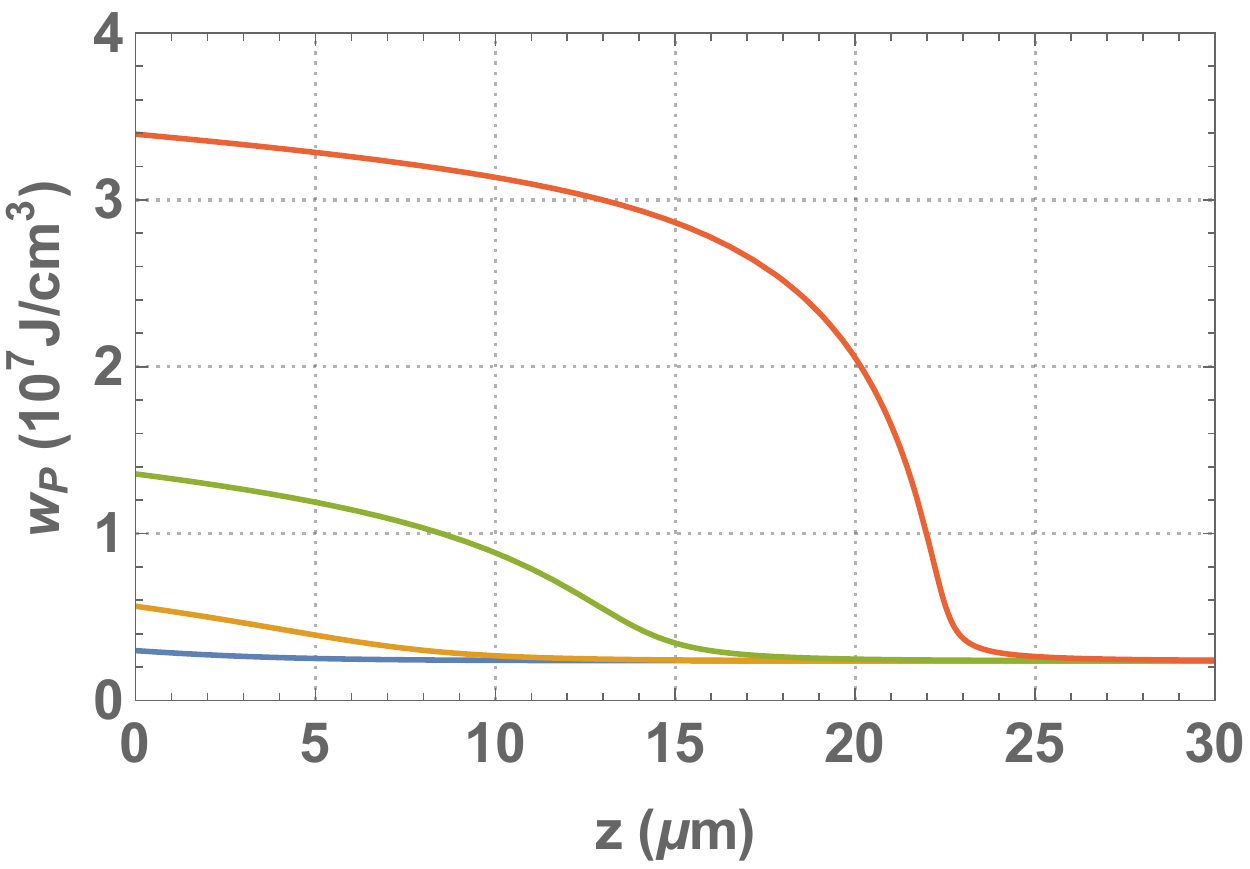}
    \caption{\label{fig:up}}
  \end{subfigure}
  \captionsetup{justification=raggedright,singlelinecheck=false}
  \caption{\label{fig:dudz} (a) Temperature of plasma at $t=10$ ps (red) and $t=1$ ns (black). (b) Energy density of plasma at t = 1 ps, 10 ps, 100 ps and 1 ns (from left to right). Parameter are the same as in fig.\ref{fig:Efield}.}
\end{figure*}

Equ.\ref{equ:heating} is solved with initial conditions same to that in our previous work \cite{li2019propagation}, that is, irradiance of incident RHCP laser is around $10^{14}$ $Wcm^{-2}$ ($a_0$ = 0.01) with wavelength of $\lambda_L=1.064$ $\mu{m}$, plasma has density of around $10^{23}$ $cm^{-3}$ (n = 100) and initial temperature of  $T_0$ = 100 eV, amplitude of magnetic field is around $10^{5}$ Tesla (B = 10). This initial condition exists in the region of $\nu \ll B$ in equ. \ref{equ: approximation}, where laser absorption decreases with plasma temperature but transmittance does not depend on it. The amplitude of electric field for reflected and transmitted laser at t = 10 ps is calculated and shown in fig.\ref{fig:Ef}. Poynting vector of laser propagating in plasma, which is normalized by that of incident laser, is shown in fig. \ref{fig:S} for various time, t = 1 ps, 10 ps, 100 ps and 1 ns, respectively. Plasma temperature at 10 ps and 1 ns is shown in fig. \ref{fig:Temp} and dimensionless energy density of heated plasma at 1 ps, 10 ps, 100 ps and 1 ns is shown in \ref{fig:up}.

It is found that as RHCP laser propagates inside, laser front slowly heats up plasma and is absorbed while laser at later time propagates in the heated plasma "freely".  Temperature of plasma at vacuum-plasma boundary reaches above 1.4 keV at 1 ns, which scales with time approximately as $T_b(keV)\approx 1.42* t_{ns}^{0.4}$ from the calculations at various time. The energy density of plasma at depth of 20 $\mu{m}$ increases to around $2*10^7$ $J/cm^3$ at 1 ns, and the conversion efficiency from laser to plasma is calculated to be as large as $\sim 40\%$ (radiant fluence of incident laser is around $10^{5}$ $J/cm^2$ and areal energy density of heated plasma is around $4*10^4$ $J/cm^2$). 

The dependence of laser propagation and plasma heating on initial parameters is discussed qualitatively below.
\begin{itemize}
\item Magnetic field: At higher magnetic field, propagation depth increases while temperature of heated plasma  decreases. 
\item Plasma density: At higher plasma density, propagation depth decreases while  temperature of heated plasma increases. 
\item Initial plasma temperature: At higher plasma temperature, propagation depth increases while temperature of heated plasma almost does not change. 
\item Laser irradiance: At higher laser irradiance, both propagation depth and temperature of heated plasma increase but the interaction will not be dominated by collisional absorption.
\item Laser wavelength: At larger wavelength of laser, propagation depth increases while temperature of heated plasma decreases greatly. It should be noted that required threshold of magnetic field decreases at larger laser wavelength. For example, $B_c$ is as "low" as 1000 tesla for powerful $CO_2$ laser, as being pursued in BESTIA project \cite{pogorelsky2016bestia}.
\end{itemize} 

The energy density of $2*10^7$ $Jcm^{-3}$ obtained here is translated to $\sim 8*10^6$ $J/g$ for specific energy, which is larger than the specfic energy of $6.4*10^6$ $J/g$ generated by accelerated heavy ions in Facility for Antiprotons and Ion Research \cite{tahir2007hedgehob}. It is also translated to pressure of $\sim200$ Mbar, which is higher than shock pressure generated by laser ablation or hot electrons from parametric instabilities of stimulated Raman scattering and two-plasmon instability in scheme of shock ignition \cite{batani2014physics}. Thus, this phenomena might find applications in  fields such as warm dense matter, high energy density plasma or inertial confinement fusion. To reach higher pressure as required by applications such as ignition, one RHCP laser and one LHCP laser could be irradiated in opposite directions in the magnetized plasma and then head-on collide, as depicted in fig. 3 in our previous work \cite{li2019propagation}. Profile of the required strong magnetic field is cylindrical, which could be generated by compression of initially much lower magnetic field. Assuming radius of cylindrical magnetic field is 20 micrometer and length of 0.2 millimeter, the energy of static magnetic field is as low as 1 kJ, since  energy density of magnetic field is around $4*10^9$ $J/cm^3$ at $10^5$ tesla. And initial plasma temperature of 100 eV could be realisticly obtained by some degree of fuel compression. By the way, this approach is, somehow, similar to scheme of fast ignition \cite{tabak2014alternative} and Magnetized Liner Inertial Fusion \cite{peterson2016demonstrating}.

The limitation of equ. \ref{equ: approximation} is also mentioned here. In the above, the initial plasma temperature is chosen to be 100 eV, where Spitzer model for collisional absorption is valid at the given plasma density of $10^{23}$ $cm^{-3}$. However, in the range of lower temperature, i.e., temperature of 10 - 100 eV, the electrons are strongly coupled and below 10 eV they are degenerate and collision frequency depends on ion temperature, thus processes other than collisional absorption should be investigated. It could be imagined that RHCP laser propagates into and heat up plasma, although with smaller depth. Furthermore, processes such as dynamic ionization and plasma hydrodynamics are not included in equ. \ref{equ: approximation}, which are applied in particle-in-cell simulation giving similar results \cite{wu2020uniform}. 
  
\section{CONCLUSION}  
Process of laser propagation and collisional heating of highly magnetized over-dense plasma at moderate laser irradiance is investigated from basic equations.  Plasma with density of $10^{23}$ $cm^{-3}$ and initial temperature of 100 eV is heated up to more than 1 keV within depth of around 20 $\mu{m}$ at 1 nanosecond by RHCP laser with irradiance of $10^{14}$ $Wcm^{-2}$, where energy density of plasma is around $2*10^7$ $J/cm^3$ and conversion efficiency from laser to plasma is as large as $40\%$. With the development of strong magnetic field, especially with field above $10^5$ tesla, volume above $(10\sim 100 \mu{m})^3$ and duration longer than 1 nanosecond, this phenomena might find some applications in the fields of high energy density plasma or even inertial confinement fusion in the future.
\\
\\
\textbf{\large{ACKNOWLEDGMENTS}}
\newline
The authors gratefully acknowledge the discussion with Prof. Vladimir Tikhonchuk of Centre Lasers Intenses et Applications and University of Bordeaux.
\\
\\
\noindent
\textbf{\large{Data availability}}.
The data that support the findings of this study are available from the corresponding author upon reasonable request.

\small
\bibliographystyle{apsrev4-1}
\bibliography{references}

\end{document}